\documentclass[conference]{IEEEtran}
\IEEEoverridecommandlockouts
\usepackage{cite}
\usepackage{amsmath,amssymb,amsfonts}
\usepackage{algorithmic}
\usepackage{graphicx}
\usepackage{textcomp}
\usepackage{xcolor}
\usepackage{hyperref}
\usepackage{booktabs}
\usepackage{array}
\usepackage{multirow}
\usepackage{enumitem}
\usepackage{balance}
\usepackage{orcidlink}
\def\BibTeX{{\rm B\kern-.05em{\sc i\kern-.025em b}\kern-.08em
    T\kern-.1667em\lower.7ex\hbox{E}\kern-.125emX}}
\begin{document}

\title{Dead Code Doesn't Talk: Authentic Requirements Elicitation in Introductory Software Engineering}

\author{\IEEEauthorblockN{Santiago Berrezueta-Guzman\,\orcidlink{0000-0001-5559-2056}}
\IEEEauthorblockA{
\textit{Technical University of Munich}\\
Heilbronn, Germany \\
}
\and
\IEEEauthorblockN{Vanesa Metaj\,\orcidlink{0009-0008-6576-346X}}
\IEEEauthorblockA{
\textit{Technical University of Munich}\\
Heilbronn, Germany \\
}
\and
\IEEEauthorblockN{Stefan Wagner\,\orcidlink{0000-0002-5256-8429}}
\IEEEauthorblockA{
\textit{Technical University of Munich}\\
Heilbronn, Germany \\
}
}

\maketitle

\begin{abstract}
Requirements elicitation is among the most communication-intensive activities in software engineering, yet it receives limited explicit treatment in undergraduate curricula. This paper presents a case study of an \emph{Introduction to Software Engineering} course in which 20 student teams applied requirements elicitation practices to a Java-based 2D game they had built in a prior programming course, engaging 18 campus doctoral and postdoctoral researchers as authentic clients.
Structured across four phases—preparation, client meeting, requirements elaboration, and a prototype sprint—the activity produced 203 elicited requirements, SRS documents with a mean quality score of $6.79 \pm 1.08$ out of 10, and prototype demonstrations scoring $7.21 \pm 1.15$. A pre/post self-assessment survey revealed statistically significant improvements across all eight measured soft-skill dimensions, with the largest gains in Stakeholder Empathy ($\Delta = +1.33$) and Negotiation ($\Delta = +1.13$). Thematic analysis of reflective reports identified four dominant learning themes, with the tension between client wishes and technical feasibility cited as the most professionally relevant experience. Our findings suggest that anchoring elicitation practice to a student-authored artifact lowers cognitive barriers while increasing authenticity, and that campus researchers serve as an accessible and effective proxy client for programs without established industry partnerships.
\end{abstract}


\section{Introduction}

The software engineering profession demands more than technical proficiency. Studies consistently show that communication failures, unclear requirements, and stakeholder misalignment remain among the leading causes of project failure~\cite{lehtinen2014perceived}. Despite this, introductory software engineering courses often treat requirements engineering as a theoretical module—taught via textbooks, followed by a simplified in-class exercise—rather than as a live, socially embedded practice \cite{daun2023systematic}.
The result is a well-documented gap: graduates can describe the phases of the Software Development Life Cycle (SDLC) but struggle to conduct a productive client interview, manage conflicting stakeholder priorities, or translate ambiguous natural-language requests into actionable specification artefacts~\cite{ouhbi2015requirements, radermacher2013gaps}.

This case study comes from a B.Sc. Information Engineering programme. In the first semester, students take Fundamentals of Programming (FoP), which uses a Project-Based Learning approach built around a 2D Maze Runner game in Java. By the end of the course, each team has a working, documented codebase covering object-oriented design, collision logic, maze loading, and a GUI.
In the next semester, the same students take Introduction to Software Engineering (ISE). Instead of starting from scratch, ISE builds directly on what they already made: teams approach real campus stakeholders, run requirements-elicitation sessions, and write formal specification documents for extensions to their own game. This creates a genuinely authentic learning context — students know their codebase inside out, clients can interact with a live prototype, and the resulting feature requests are both technically grounded and user-driven.

This paper makes the following contributions: (1) A replicable pedagogical design for embedding client-facing requirements elicitation into an introductory SE course using a pre-existing student artefact. (2) A structured analysis of how diverse client backgrounds (non-technical vs.\ technical) influence the complexity and realism of elicited requirements. (3) A scalable roadmap applicable to other SE courses that wish to integrate authentic stakeholder interaction.

\section{Related Work}\label{sec:related}

Requirements engineering (RE) education has long been identified as underserved in undergraduate curricula \cite{soo2018game}. 
The use of real clients in educational projects has been shown to increase student motivation and strengthen connections between academic learning and professional practice across multiple disciplines~\cite{parsons2009group}. 

Steghöfer et al.~\cite{steghofer2018involving} conducted a comprehensive study across eight software engineering project courses at multiple universities, developing a conceptual model for planning and evaluating stakeholder involvement. They confirmed that external stakeholders improve student motivation, but also reveal some problems, such as misalignment between stakeholder expectations and student capabilities, inconsistent engagement, and scheduling incompatibilities. Thus, they confirm that this method has a significant impact on course outcomes only when it is properly monitored.

Even though the software engineering industry increasingly prioritizes interpersonal abilities, with research showing that up to 75\% of long-term job success depends on soft skills \cite{article}, traditional computing curricula often struggle to integrate them effectively, with soft skills frequently delegated to generic university communication courses that fail to address the specific challenge of translating technical constraints for non-technical stakeholders \cite{10.1145/1953163.1953312}. To bridge this gap, educators have turned to experiential and project-based learning. Recent studies demonstrate that situating soft skill development within student-centered technical projects produces improvements in active listening, teamwork, and stakeholder empathy \cite{article, motschnig2026long}. 

Project-Based Learning (PBL) has become a highly effective pedagogical strategy in computer science to encourage problem-solving and critical thinking ~\cite{rehman2023trends}, with game development proving particularly effective at generating authentic motivation through creative authorship and personal ownership of code \cite{diaz2026gamified}.
Teaching Agile in academic settings is challenging, as students often treat it as administrative overhead rather than a practical necessity \cite{devedvzic2010teaching}. However, when integrated with real stakeholder reviews and iterative deliveries, students transition from mere compliance to genuine process adoption \cite{sa2025bridging}.

\section{Course and Activity Design}\label{sec:design}

\subsection{Course Context}

\emph{Introduction to Software Engineering} (ISE) is a second-semester mandatory course for the B.Sc. Information Engineering students. It follows directly from \emph{Fundamentals of Programming (FoP)} and assumes that students have working knowledge of Java, object-oriented design, and collaborative development. ISE covers the full SDLC with emphasis on: (i) agile and plan-driven process models; (ii) requirements engineering; (iii) architectural and detailed design; (iv) testing strategies; and (v) project management fundamentals. The course runs for 12 weeks with two lecture hours and a two-hour lab session per week.

\subsection{Student Demographics and Prerequisites}
The course consisted of 135 students, approximately 75\% male and 25\% female. Students represented 30 different nationalities, with the largest groups originating from East Asia, Southeast Asia, and the Middle East.

All participating students were required to have completed the predecessor course \emph{Fundamentals of Programming}, ensuring sufficient coding proficiency to shift focus toward soft skills and client-facing elicitation.


\subsection{The Predecessor Artefact: The Maze Runner Game}

The software artefact is a \emph{2D Maze Runner} style game, originally developed by the student teams during their prerequisite first-semester \emph{Fundamentals of Programming (FoP)} course.
This artefact was deliberately chosen as the anchor for the \emph{Introduction to Software Engineering} (ISE) elicitation exercise because it satisfies two critical conditions. First, it features an interactive product that non-technical stakeholders can easily understand and play with during live demonstrations, which sparks feature requests. Second, the students authored the codebase themselves and are familiar with the system's architectural constraints. This technical ownership empowers students to immediately evaluate the feasibility of a client's request during the interview, arriving at realistic negotiations.

\subsection{Client Recruitment and Profile}

Clients were doctoral and postdoctoral researchers across all chairs in the Computer Science Faculty, selected because they are easily accessible on campus, span a wide range of technical literacy levels mirroring real-world client heterogeneity, and are accustomed to analytical thinking that produces structured feature requests rather than vague wishes. 

Participation was voluntary. Clients were asked to attend one 15-minute meeting, interact with the running game, express genuine feature wishes, and be available for at most one follow-up clarification by email. No compensation beyond the reward of influencing a student project was provided.



\subsection{Activity Structure and Timeline}

The elicitation activity was structured as a four-phase sequence integrated into the semester, illustrated by Figure \ref{Phases}:

\begin{figure*}[h!]
    \centering
    \includegraphics[width=0.9\linewidth]{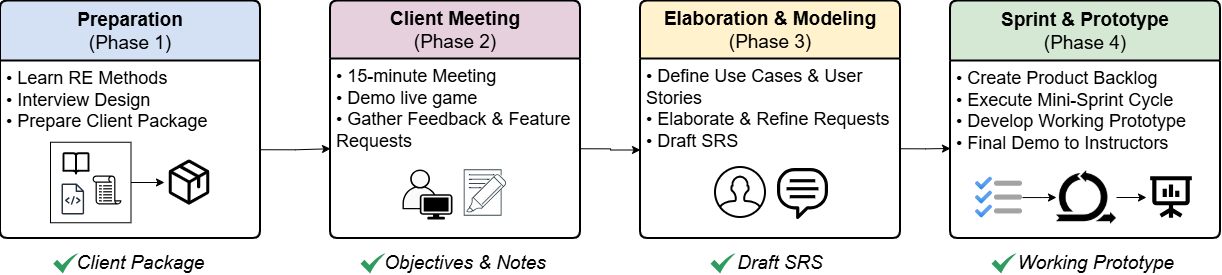}
    \caption{Four Phases of ISE Requirements Activity}
    \label{Phases}
\end{figure*}

\paragraph{Phase 1 — Preparation}

Students reviewed RE theory (elicitation techniques, use-case modelling, user stories) through lectures and a structured reading list. Lab sessions covered interview design: how to prepare open-ended questions, avoid leading prompts, and manage a meeting agenda. Each team prepared a \emph{client package} consisting of a one-page game overview, a demonstration script, and a blank requirements-capture template.

\paragraph{Phase 2 — Client Meeting}
Teams conducted a single, 15-minute in-person meeting with their assigned client (one client per team), and each client could be assigned to up to 2 teams to introduce conflicting priorities. Meetings took place wherever the clients decided, on-site at the campus or online. It was instructed that the teams must organize their contributions in the meeting; therefore, one team member was suggested to run the game demonstration and facilitate the conversation; another took structured notes; and the third managed the time and ensured all agenda items were covered. Meetings were not audio-recorded for later instructor review.

\paragraph{Phase 3 — Requirements Elaboration}
Teams processed their raw notes into formal artefacts: a user story map, a use-case diagram, and a draft Software Requirements Specification (SRS) following the IEEE~830 template. During this phase, students were permitted one asynchronous follow-up exchange with their client to clarify ambiguities. 

\paragraph{Phase 4 — Process Adaptation and Prototype Sprint}

With an approved SRS, teams entered a mini-sprint cycle. They decomposed requirements into a product backlog, estimated effort using story points, and planned a two-sprint development cycle targeting a working prototype of at least two of the elicited features. Weekly lab check-ins replaced the sprint retrospective. The final deliverable was a live demonstration to instructors and, where available, to the original client pair.

\section{Assessment Methodology}\label{sec:methodology}

\subsection{Assessment Dimensions}

Assessment in ISE for this activity was structured across four complementary dimensions, mirroring the multi-modal philosophy of the predecessor FoP course~\cite{berrezueta2026beyond}:

\textbf{D1. Requirements Artefact Quality} — Evaluated the completeness, consistency, and traceability of the SRS, user story map, and use-case diagram. Graded using an adaptation of the RE quality framework~\cite{dos2020software}.

\textbf{D2. Client Meeting Performance} — Assessed via an instructor survey of the clients to understand how well the meeting facilitation, question quality, active listening, and time management were.

\textbf{D3. Prototype Demonstration} — Evaluated the fidelity of the implemented new prototype to the elicited new requirements, architectural soundness, and the ability to justify design decisions under questioning.

\textbf{D4. Reflective Report and Peer Assessment} — Each student submitted an individual reflection on their personal contribution and their observations of the team's process adaptation to the new client's requirements.

\subsection{Grading Rubric and Weight Distribution}

The overall activity accounted for 50\% of the ISE course grade,
divided as shown in Table~\ref{tab:rubric}.

\begin{table}[htbp]
\caption{Assessment Weight Distribution}
\label{tab:rubric}
\centering
\renewcommand{\arraystretch}{1.3}
\begin{tabular}{p{0.45\columnwidth} c c}
\toprule
\textbf{Assessment Dimension} & \textbf{Weight (\%)} & \textbf{Format} \\
\midrule
Requirements Artefact Quality  & 30 & Team \\
Client Meeting Performance     & 30 & Team \\
Prototype Demonstration        & 30 & Team \\
Individual Reflection          & 10 & Individual \\
\midrule
\textbf{Total}                 & \textbf{100} & \\
\bottomrule
\end{tabular}
\end{table}

\subsection{Soft-Skills Evaluation Instrument}

Assessing soft skills reliably is an established challenge in computing education~\cite{andree2025soft}. In this study, soft-skill development was captured through two instruments:

\textbf{Pre/post self-assessment survey}: Students rated their perceived competence across eight dimensions (communication, active listening, negotiation, conflict resolution, documentation, time management, adaptability, and stakeholder empathy) on a five-point Likert scale at the start of ISE and again at the end of the elicitation activity.

\textbf{Client/Instructor observation rubric}: Applied during the client meeting review and the prototype demonstration to provide an external perspective on communication quality. This is done by analyzing the clients who completed a brief post-meeting survey rating the students' professionalism, preparedness, soft skills, and overall meeting quality.

\subsection{Data Collection and Analysis}

This study adopted a mixed-methods design. Quantitative data were collected from three sources: (i) the pre-/post-soft-skill self-assessment surveys; (ii) the SRS grading rubric scores applied by two independent instructors; and (iii) the post-meeting client satisfaction surveys. Because Likert-scale responses are ordinal, pre-/post-differences were analysed using the Wilcoxon signed-rank test with Cliff's $\delta$ as the effect-size estimator. SRS dimension scores were treated as interval-level data and summarised using means and standard deviations. Inter-rater reliability for the SRS grading rubric was established across a stratified 25\% sample of documents and assessed using Cohen's $\kappa$.

Qualitative data from the individual reflective reports ($N = 60$) were coded thematically following the six-phase procedure described by Braun and Clarke~\cite{braun2006using}: familiarisation, generating initial codes, searching for themes, reviewing themes, defining and naming themes, and writing up. Two researchers coded independently, resolving disagreements through discussion until consensus was reached. Final inter-coder reliability was $\kappa = 0.78$, indicating substantial agreement.

\section{Results and Analysis}\label{sec:results}

\subsection{Client Meeting Outcomes}

The cohort comprised 60 students organized into 20 teams of three. A total of 18 campus clients were recruited (14 doctoral students and 4 postdoctoral researchers), distributed across the CIT school. All 20 meetings were held within the established timeslot, and no team missed their meeting.

Table~\ref{tab:client_satisfaction} summarizes client satisfaction. Overall meeting quality was rated at $3.74 \pm 0.65$, indicating a solid but not exceptional first client-interaction experience. Professionalism received the highest rating ($3.82 \pm 0.71$), reflecting the preparation effort invested in Phase~1. Active listening received the lowest score ($3.54 \pm 0.88$), a finding consistent with the instructor observation rubric, which noted a tendency for students to redirect conversations toward technical topics before fully exhausting the client's expressed needs. Open-ended client comments praised the live game demonstration as a strong engagement tool; several clients noted it prompted them to articulate feature wishes they had not previously formed.

\begin{table}[htbp]
\caption{Client Satisfaction Survey Results (Mean $\pm$ SD, scale 1–5)}
\label{tab:client_satisfaction}
\centering
\renewcommand{\arraystretch}{1.3}
\begin{tabular}{lcc}
\toprule
\textbf{Dimension} & \textbf{Mean} & \textbf{SD} \\
\midrule
Professionalism          & 3.82 & 0.71 \\
Preparedness             & 3.61 & 0.83 \\
Active Listening         & 3.54 & 0.88 \\
Overall Meeting Quality  & 3.74 & 0.65 \\
\bottomrule
\end{tabular}
\end{table}

Across all 20 SRS submissions, 203 distinct requirements were identified and categorised. Table~\ref{tab:req_types} shows the full distribution. Gameplay Mechanics Extensions constituted the largest category (25.6\%), followed by UI/UX Improvements (20.2\%) and Narrative/Thematic Additions (15.8\%). Together, these three surface-level experiential categories account for over 60\% of all requirements, reflecting the client pool's predominantly non-technical composition.

Notably, the two clients assigned to two teams each generated measurably divergent requirement sets when interviewed separately, confirming that the shared-client design successfully introduced conflicting priorities.

\begin{table}[htbp]
\caption{Distribution of Elicited Requirement Types}
\label{tab:req_types}
\centering
\renewcommand{\arraystretch}{1.3}
\begin{tabular}{lcc}
\toprule
\textbf{Requirement Category} & \textbf{Count} & \textbf{\%} \\
\midrule
Gameplay Mechanics Extensions   & 52  & 25.6\% \\
UI / UX Improvements            & 41  & 20.2\% \\
Narrative / Thematic Additions  & 32  & 15.8\% \\
AI / Algorithmic Features       & 28  & 13.8\% \\
Accessibility Features          & 22  & 10.8\% \\
Performance / Technical NFRs    & 17  &  8.4\% \\
Multiplayer / Networking        & 11  &  5.4\% \\
\midrule
\textbf{Total}                  & \textbf{203} & \textbf{100\%} \\
\bottomrule
\end{tabular}
\end{table}

\subsection{Requirements Artefact Quality}

SRS documents were evaluated across four quality dimensions using the RE quality framework~\cite{hickey2004unified}: completeness, consistency, clarity, and traceability. Results are shown in Table~\ref{tab:srs_quality}. The overall mean SRS score was $6.79\pm 1.08$ out of 10, indicating a competent first attempt at formal specification for a cohort with no prior RE experience.

\begin{table}[htbp]
\caption{SRS Quality Scores (Mean $\pm$ SD, scale 0–10)}
\label{tab:srs_quality}
\centering
\renewcommand{\arraystretch}{}
\begin{tabular}{lcc}
\toprule
\textbf{Quality Dimension} & \textbf{Mean} & \textbf{SD} \\
\midrule
Completeness    & 6.83 & 1.42 \\
Consistency     & 7.14 & 1.18 \\
Clarity         & 7.31 & 1.09 \\
Traceability    & 5.87 & 1.63 \\
\midrule
\textbf{Overall SRS Score} & 6.79 & 1.08 \\
\bottomrule
\end{tabular}
\end{table}

Clarity was the strongest dimension ($7.31 \pm 1.09$). Post-hoc analysis suggested that this was attributable to students writing for an audience they had personally met: the client's vocabulary, background, and reactions to the live demo served as fresh reference points when drafting requirement statements. Consistency was similarly strong ($7.14 \pm 1.18$), likely aided by the shared use of the IEEE 830 template. 

Traceability was the weakest dimension ($5.87 \pm 1.63$) and showed the highest variance, indicating wide differences in students' understanding of the concept. While most teams produced a user story map and a use-case diagram, fewer than half explicitly linked these artefacts to specific SRS requirement identifiers or traced requirements forward to acceptance criteria. Indeed, non-functional requirements (NFRs) accounted for only 8.4\% of all elicited requirements (Table~\ref{tab:req_types}), underscoring that functional features dominate novice elicitation sessions at the expense of operational quality attributes.
Inter-rater reliability for the SRS rubric across the stratified sample was $\kappa = 0.74$ (substantial agreement), confirming that the grading rubric was sufficiently operationalised for consistent application across assessors.

\subsection{Soft-Skills Development}

\subsubsection{Pre/Post Self-Assessment}

Table~\ref{tab:softskills} reports the mean self-assessed competence scores across all eight dimensions at the start of ISE (Pre) and immediately after the completion of Phase~3 (Post). All eight dimensions showed positive mean shifts. Seven reached statistical significance at $p < 0.05$ under the Wilcoxon signed-rank test; Time Management was the sole exception ($p = 0.031$, which nonetheless crosses the threshold, but with the smallest effect size, Cliff's $\delta = 0.24$).

\begin{table}[htbp]
\caption{Soft-Skill Self-Assessment: Pre vs.\ Post Scores (Mean $\pm$ SD)}
\label{tab:softskills}
\centering
\renewcommand{\arraystretch}{1.3}
\begin{tabular}{lcccc}
\toprule
\textbf{Skill Dimension} & \textbf{Pre} & \textbf{Post} & \textbf{$\Delta$} & \textbf{$p$} \\
\midrule
Communication       & $2.93 \pm 0.81$ & $3.72 \pm 0.74$ & $+0.79$ & $<0.001$ \\
Active Listening    & $3.12 \pm 0.76$ & $3.87 \pm 0.65$ & $+0.75$ & $<0.001$ \\
Negotiation         & $2.28 \pm 0.91$ & $3.41 \pm 0.83$ & $+1.13$ & $<0.001$ \\
Conflict Resolution & $2.51 \pm 0.85$ & $3.18 \pm 0.79$ & $+0.67$ & $0.003$  \\
Documentation       & $3.05 \pm 0.72$ & $3.83 \pm 0.61$ & $+0.78$ & $<0.001$ \\
Time Management     & $3.21 \pm 0.79$ & $3.58 \pm 0.73$ & $+0.37$ & $0.031$  \\
Adaptability        & $2.84 \pm 0.83$ & $3.71 \pm 0.70$ & $+0.87$ & $<0.001$ \\
Stakeholder Empathy & $2.19 \pm 0.94$ & $3.52 \pm 0.82$ & $+1.33$ & $<0.001$ \\
\bottomrule
\end{tabular}
\end{table}

The two largest gains were in Stakeholder Empathy ($\Delta = +1.33$, Cliff's $\delta = 0.68$, large effect) and Negotiation ($\Delta = +1.13$, Cliff's $\delta = 0.59$, large effect). These are precisely the skills that are most directly exercised when a student must understand and then respectfully push back on a real client's feature request. The fact that both dimensions began at the lowest pre-scores ($2.19$ and $2.28$ respectively) further indicates that, prior to the activity, students had little prior exposure to these competencies in a professional context—consistent with the cohort's first-semester background.

Communication and Documentation both showed substantial gains ($\Delta = +0.79$ and $+0.78$). The improvement in Documentation is particularly noteworthy, given that FoP already required Javadoc commenting; the higher post-scores suggest that writing for a human stakeholder audience (as required by the SRS) develops a qualitatively different and more demanding form of documentation competency.

Time Management showed the smallest gain ($\Delta = +0.37$), which is consistent with the fact that students had already developed scheduling habits through the FoP project milestones in the prior semester. The ceiling effect for this dimension (high pre-score of $3.21$) further limits the observable delta.

\subsubsection{Qualitative Themes from Reflective Reports}

Thematic analysis of the individual reflective reports surfaced four dominant themes:

\textbf{T1. Bridging technical and non-technical discourse.}
The majority of students noted the challenge of translating technical constraints (e.g., the rendering pipeline's limitations, the absence of a networking layer) into language accessible to a non-technical client, and conversely of parsing informal client wishes into unambiguous requirement statements. Several students reported that writing the SRS forced them to revisit and refine their mental model of the game's own architecture.

\textbf{T2. Ownership and pride in the existing artefact.}
Students repeatedly commented that presenting \emph{their own} game—rather than a given specification—created a sense of personal stake in the meeting. A representative reflection stated: \textit{``It felt different to show something we actually built. We wanted to impress the client, so we prepared more carefully than for any lab session.''} This intrinsic motivation was not directed by the grading rubric and suggests that artefact ownership is a genuine driver of preparation effort.

\textbf{T3. Tension between client wishes and technical feasibility.}
Teams consistently reported moments of negotiation in which a client's request was deemed too complex or architecturally disruptive to the sprint timeline. Students described the challenge of communicating ``no—but perhaps in a future sprint'' without appearing dismissive, an experience they identified as novel and professionally relevant.

\textbf{T4. Process adaptation as a learning moment.}
The transition from the FoP development style—informal task distribution, no explicit backlog—to a story-point-estimated, sprint-structured workflow was frequently cited as cognitively demanding but ultimately satisfying, 
with the external commitment to a real client making abstract agile concepts concrete. 

\section{Discussion}\label{sec:discussion}
Using a codebase students have already built differs markedly from providing a fictional system description for a requirements exercise. Artefact familiarity produced two distinct benefits: \textbf{reduced cognitive load}, as students did not need to simultaneously learn a new domain and practise elicitation techniques, freeing cognitive resources for the interpersonal dimensions of the meeting, and \textbf{grounded feasibility reasoning}, as their intimate knowledge of the codebase allowed immediate, concrete feasibility assessments during the interview rather than deferring judgment to a analysis phase.




\subsection{Team Process Adaptation}

The mean prototype score of $7.21 \pm 1.15$ and the $70\%$ rate of teams exceeding the minimum feature count indicate that the transition from informal FoP-style development to sprint-structured delivery was largely successful. The two teams that implemented only the minimum features both provided explanatory rationales in their reflective reports: one attributed it to a mid-sprint architectural pivot driven by a client clarification; the other to underestimation during backlog planning. Both are authentic project management failure modes that generated richer reflective reports than teams whose sprints ran smoothly.
The reflective data in Theme~T4 suggest that the source of sprint discipline was not the instructor-mandated process but the moral commitment to a real client. This finding aligns with the broader agile education literature, which identifies external accountability as one of the most effective catalysts for process adoption in student teams~\cite{devedvzic2010teaching}.

\subsection{Instructor and Tutor Role}
Instructors and tutors operated as facilitators. During Phase 1, tutors reviewed each team's client package and interview guide, providing formative feedback on the quality of the questions. During Phases 3 and 4, weekly stand-up check-ins replaced formal retrospectives, with tutors probing prioritisation decisions without prescribing solutions. This choice is consistent with the PBL literature's recommendation to maintain learner agency \cite{rehman2023trends} and avoid over-involvement that suppresses authentic problem-solving.


\subsection{Observed Student Patterns}
Analysis of client satisfaction data and reflective reports identified three recurring behavioural patterns. First, role specialisation solidified rapidly: in nearly all teams, the same student who had led the FoP presentation assumed the facilitator role in the client meeting, suggesting that communication leadership roles persist across course transitions. Second, overuse of technical jargon in early exchanges was the most common trigger for low Active Listening scores; client comments frequently noted being ``talked past'' during the first five minutes before students settled into a more responsive conversational mode. Third, several teams initially resisted requirements outside their perceived technical comfort zone, particularly accessibility features and narrative additions.

\subsection{Scalability and Adaptability}
The pedagogical model is transferable to other SE courses with three minimal conditions: (a) students possess a runnable, documented software artefact they have personally authored; (b) a pool of non-student stakeholders is accessible on or near campus; and (c) the course timeline accommodates the four-phase sequence. The model does not require industry partnerships — a significant advantage for programmes without established corporate connections.


\subsection{Limitations}

\textbf{Single-course context.} All data derive from one cohort at one institution. Replication across diverse institutional contexts is required to assess generalisability.

\textbf{Self-selection of clients.} Campus researchers who volunteer may be more engaged and constructive than a random industrial sample. Client satisfaction scores may therefore be optimistically biased upward.

\textbf{Self-report validity.} Soft-skill self-assessments are subject to social desirability bias and response-shift bias; students may have reassessed their pre-activity competence level retrospectively. The instructor observation rubric provides a partial external signal but was available only for meeting performance, not for all eight dimensions.

\textbf{Absence of a control group.} A within-cohort control condition, ethically challenging to implement, would be needed to isolate the causal contribution of the elicitation activity.

\section{Conclusions and Future Work}\label{sec:conclusion}
This paper reports a client-facing requirements elicitation activity embedded in a second-semester \emph{Introduction to Software Engineering course}, in which 60 students across 20 teams engaged 18 campus researchers as authentic clients, using their own Java Maze Runner game as the elicitation anchor.
The activity produced 203 distinct requirements spanning seven categories, with experiential features dominating over algorithmic ones. SRS clarity was the strongest quality dimension ($7.31 \pm 1.09$) while traceability was the weakest ($5.87 \pm 1.63$), consistent with the RE education literature on novice specifiers. All eight soft-skill dimensions improved significantly, with Stakeholder Empathy ($\Delta = +1.33$) and Negotiation ($\Delta = +1.13$) showing the largest gains — precisely the competencies most directly exercised by live client interaction.


Collectively, these results support the conclusion that a structured, artifact-anchored elicitation activity can meaningfully develop the communicative, analytical, and process management skills most frequently lacking in new graduates, even within a single undergraduate semester. 
Educators replicating this model are encouraged to follow these steps: (i) identify a visually demonstrable student-authored artefact; (ii) recruit a diverse client pool spanning technical and non-technical backgrounds; (iii) structure the activity into at least four phases with explicit preparation scaffolding; and (iv) close the loop with a client-attended final demonstration, which students consistently identified as the moment of highest perceived professional relevance.

Future work will replicate the activity with industry clients made available through the institution's technology transfer office, to determine whether commercial stakeholders generate qualitatively different requirement types and whether the higher-stakes context further elevates student motivation and preparation effort.

\section*{Acknowledgements}

The authors thank the doctoral students and postdoctoral researchers
who volunteered their time as project clients.

\bibliographystyle{ieeetr}
\bibliography{00References}
\end{document}